\documentclass[runningheads]{llncs}

\usepackage{amsfonts}
\usepackage{amssymb}
\usepackage{graphicx}
\usepackage[utf8x]{inputenc}
\usepackage[ruled]{algorithm2e}
\usepackage{amsmath}

\begin{document}

\title{A Taxonomy of the Biases of the Images created by Generative Artificial Intelligence}
\titlerunning{Taxonomy of Biases of Image Generative AI}

\author{Adriana Fernández de Caleya Vázquez, Eduardo C. Garrido-Merchán}
\date{March 2023}

\institute{Universidad Pontificia Comillas, Madrid, Spain \\
\email{adricaleya@alu.comillas.edu, ecgarrido@icade.comillas.edu}}

\maketitle 

\abstract{Generative artificial intelligence models show an amazing performance creating unique content automatically just by being given a prompt by the user, which is revolutionizing several fields such as marketing and design. Not only are there models whose generated output belongs to the text format but we also find models that are able to automatically generate high quality genuine images and videos given a prompt. Although the performance in image creation seems impressive, it is necessary to slowly assess the content that these models are generating, as the users are uploading massively this material on the internet. Critically, it is important to remark that generative AI are statistical models whose parameter values are estimated given algorithms that maximize the likelihood of the parameters given an image dataset. Consequently, if the image dataset is biased towards certain values for vulnerable variables such as gender or skin color, we might find that the generated content of these models can be harmful for certain groups of people. By generating this content and being uploaded into the internet by users, these biases are perpetuating harmful stereotypes for vulnerable groups, polarizing social vision about, for example, what beauty or disability is and means. In this work, we analyze in detail how the generated content by these models can be strongly biased with respect to a plethora of variables, which we organize into a new image generative AI taxonomy. We also discuss the social, political and economical implications of these biases and possible ways to mitigate them.}

\keywords{Generative AI; Image Generation; Fairness; AI Ethics}

\section{Introduction}

Artificial Intelligence (AI) has significantly advanced \cite{castro2016promise}, particularly in the area of image generation, where it stands out for its ability to produce illustrations that are both detailed and realistic \cite{goring2023analysis}. This progress emphasizes AI's growing influence across various applications, from enhancing facial recognition technologies to developing digital content creation \cite{chauhan2024identifying}. As AI grows, it is being integrated into different aspects of everyday life, but at the same time, ethical challenges about its performance are also increasing. An alarming aspect of the possible biases that may exist in the generation of images is the perpetuation of stereotypes and moral damage to vulnerable groups. An illustrative study found that 95\% of images generated by the popular model Stable Diffusion from Stability AI for the prompt "playing basketball" predominantly featured African American men, therefore highlighting the risk of such technologies amplifying racial and gender stereotypes \cite{chauhan2024identifying}. 

These biases originate from multiple sources \cite{ferrara2023fairness}, including the datasets used to train AI image generation models, the design of the algorithms, and the interpretation of the outputs by end-users \cite{tan2020improving}. These biases can have harmful effects on minority groups, further propagating social inequalities. Consequently, there is an urgent need to identify, comprehend, and mitigate biases in AI image generation, with the aim of achieving models that operate in a fair and inclusive manner. Previous studies in this area will serve as the basis for our research, where different types of biases within AI have been analyzed, including those related to race, gender and cultural background, as well as the social impact of these prejudices \cite{tan2020improving}. 

Identifying biases in AI image generation is challenging, as it is necessary to collect balanced and non-discriminatory training bases and to develop algorithms that are capable of detecting and rectifying biases \cite{tan2020improving}. This paper seeks to contribute to the ongoing discourse on biases within AI, specifically focusing on image generation, by proposing a classification of these biases. Moreover, this document aims to function as a practical checklist for developers of AI image generation models, guiding them to avoid these biases, therefore facilitating the creation of image generation models that embraces fairness and equity. 

This article is structured into several sections, beginning with this introduction to the issues of biases in AI image generation, followed by an in-depth analysis of these biases and their implications. It will be continued with a third technical section dedicated to describing the technical intuition behind why image generative AI models content is biased, which is necessary to understand, as well as the biases introduced in the previous section, for the discussion section. Without a technical understanding of image generative AI it is not possible to understand why the biases emerge and how can they be mitigated, this is why we emphasize in the importance of this section. After this, a discussion section will be included, where related works in the field will also be reviewed, where the social, economical and political consequences of biases in current image generative AI will be illustrated. The article concludes with insights on the wider significance of this research, contemplating future directions for investigation and the development of image generative AI methodologies that mitigate the described biases. 

\section{Taxonomy of Biases in Generative AI images}

\begin{figure}[h]
  \centering
  \includegraphics[width=1.02\textwidth]{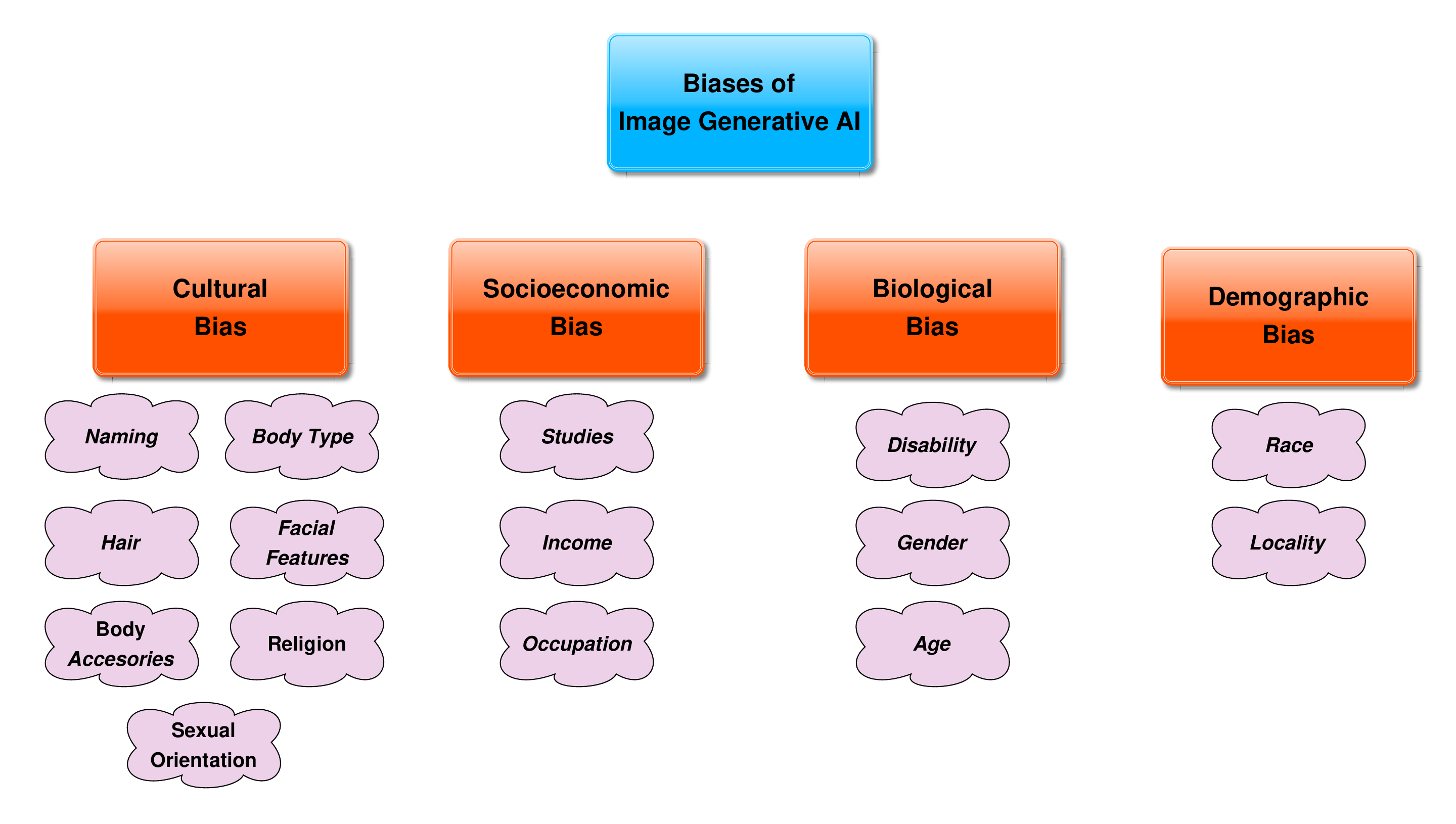} 
  \caption{Taxonomy of image generative artificial intelligence models biases}
  \label{fig:taxonomy}
\end{figure}

This section will develop a taxonomy of existing biases in AI image generation, establishing the necessary framework for understanding how these stereotypes can influence the various applications of this technology. This crucial part of the study highlights the importance of categorizing and analyzing in detail the different types of biases, from cultural to biological, to identify what their possible origins may be and how they manifest themselves. The following is a description of each category of bias included in the taxonomy that can be seen in Figure \ref{fig:taxonomy}.

\subsection{Cultural Biases}
Cultural biases refer to the biases, often unconscious biases, that favor or discriminate against certain cultures, practices, or groups based on norms, values, or beliefs prevalent in other cultures \cite{ocampo2021sesgo}. These biases manifest themselves in various ways and can significantly influence how AI modeling algorithms perceive, interpret, and represent people from different cultural backgrounds. The origin of cultural biases often comes from the datasets used to train AI models \cite{mehrabi2021survey}. If these data contain a disproportionate representation of certain cultural groups over others, or if they reflect specific cultural stereotypes, AI models can learn and perpetuate these biases. This is because AI develops patterns and associations based on the information available during its training, without questioning the fairness of these data.

\subsubsection{Naming Bias}
Name-based image generation reveals a dimension of bias where algorithms make subconscious associations between specific proper names and a set of cultural attributes, which can be perpetuated by AI algorithms. This form of bias manifests itself when an AI model produces images that reflect cultural stereotypes or expectations based on the name entered as input. 

\begin{figure}[h]
  \centering
  \includegraphics[width=0.44\textwidth]{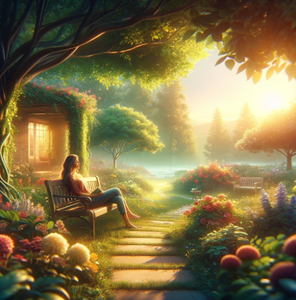}
  \includegraphics[width=0.49\textwidth]{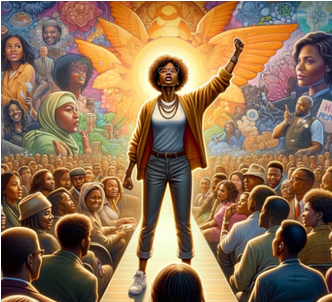}
  \caption{Naming bias example. The figure at the left is an example of image prompted with the name Laura, the figure at the right is an example of the image prompted with the name Rigoberta.}
  \label{fig:naming_bias}
\end{figure}

Unlike more commonly identified and studied biases, such as those associated with gender, race, or age, name bias covers a dimension of prejudice that can influence perceptions of cultural identities. For example, the way in which an AI model might represent a name commonly associated with a certain region or culture might be defined by unfounded assumptions. We can see an example of such a bias in Figure \ref{fig:naming_bias}.

\subsubsection{Body Type Bias}
Body type bias in AI image generation reflects an issue where certain bodies are favorably represented, perpetuating specific aesthetic ideals and making others invisible. This bias manifest itself through the preference of body types that align with dominant cultural norms in the datasets used to train AI models. Such norms often favor thin, athletic figures, neglecting a balanced representation of the existing body diversity in society. 

\begin{figure}[htb!]
  \centering
  \includegraphics[width=0.99\textwidth]{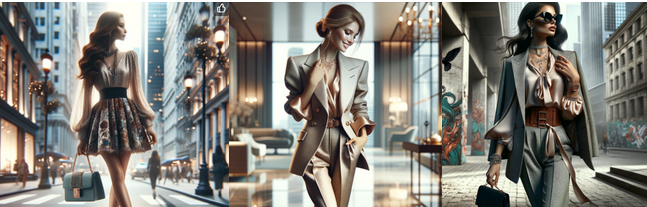}
  \caption{Body type bias. The women images generated by AI are all slim, wear elegant clothes and have white skin.}
  \label{fig:bodytype_bias}
\end{figure}

A revealing study \cite{llorente2022canones} explores the profound influence of Instagram on young women's body self-perception. The research provides evidence on how beauty canons, amplified by this platform, contribute to a distorted perception of one's own body, fomenting dissatisfaction and mental discomfort. The constant interaction with this type of content not only shapes a misconception of beauty but also drives users towards external practices in search of acceptance and social validation. Similarly, it is imminent to address the worrying spread of unattainable "perfect bodies" in AI image generation models, as these platforms have an even greater reach than this social network. We can see an example of such a bias in Figure \ref{fig:bodytype_bias}.

\subsubsection{Facial Feature Bias}
Facial feature bias highlights how perceptions and cultural norms of beauty influence the representation of individuals in digital media. This bias is evidenced by the tendency of AI models to generate images of "beautiful people" that favor features such as facial symmetry, thin faces, and prominent lips, stereotypes dictated by dominant cultures. 

AI models, when trained with data sets that lack diversity, tend to replicate and reinforce the conceptions of beauty prevalent in those sets. This practice contributes to the perpetuation of beauty ideals, excluding other expressions of facial beauty that differ from these universal criteria. An analysis on beauty stereotypes and discrimination \cite{villanueva2022estereotipos} highlights how social structures and the media perpetuate aesthetic norms, which encourages exclusion and discrimination based on appearance. The research emphasizes the urgent need to revise our ideals of beauty and adapt AI technologies to reflect the real diversity of society. We can see an example of such a bias in Figure \ref{fig:beauty_bias}.

\begin{figure}[htb!]
  \centering
  \includegraphics[width=0.99\textwidth]{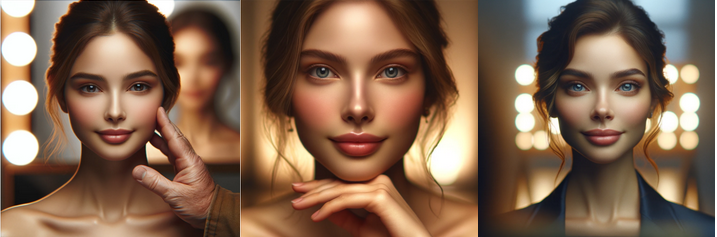}
  \caption{Facial feature bias. We can see similar patterns in the faces illustrated, sharing the same cultural beauty stereotype.}
  \label{fig:beauty_bias}
\end{figure}

\subsubsection{Hair Bias}
Hair bias brings to light the influence of cultural stereotypes on the representation of hair, demonstrating a preference for certain hair types, styles and colors that align with conventional ideals of beauty. This bias not only limits diversity in hair representation but also contributes to reinforcing inequalities and discrimination, negatively affecting the self-esteem and identity of people whose hair characteristics differ from these norms.  

Research on the impact of beauty norms in the cosmetics industry \cite{ferrara2023fairness}, suggests that the recognition and inclusion of greater hair diversity can have empowering feelings on individuals, challenging traditional stereotypes and promoting a more inclusive perception of beauty. The growing awareness and acceptance of all hair types, especially those historically excluded or stigmatized, underscore the critical importance of developing AI technologies that reflect and respect human diversity in its entirety. We can see an example of such a bias in Figure \ref{fig:beauty_bias}.

\begin{figure}[htb!]
  \centering
  \includegraphics[width=0.99\textwidth]{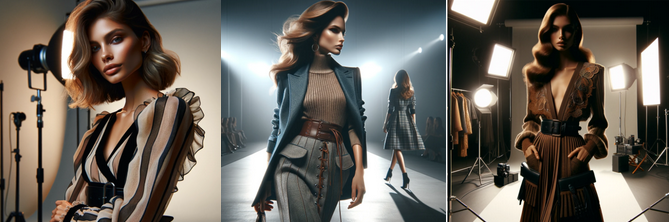}
  \caption{Hair bias. Long hair for women is always preferred.}
  \label{fig:beauty_bias}
\end{figure}

\subsubsection{Bias toward Body Accessories and Corporal Modifications}
Expressions, such as tattoos, piercings, and other body modifications, are represented or, more frequently, ignored. This bias manifests through the tendency of omitting or misinterpreting these elements, favoring representations that conform to more conventional aesthetic norms. Such omission not only limits the diversity and authenticity of human representations in digital media, but also reflects and potentially reinforces existing biases against individuals who choose these forms of expression. 

More research illuminates how body modifications can negatively influence the public perception of individuals \cite{duarte2020discriminacion}, especially in professional contexts, where they often suffer discrimination. This study highlights the real consequences of stereotypes associated with body modifications, linking them to perceptions of professional incompetence or the attribution of unfounded negative characteristics. By ignoring or inappropriately representing these forms of expression in image generation, AI not only fails to reflect positive diversity but may also contribute to the perpetuation of these discriminatory dynamics. 

\subsubsection{Religious Belief Bias}
Bias towards religiosity points out a critical aspect of how religious practices and symbols are visually displayed, where specific characteristics are often amplified to the point of creating stereotypical and overly caricatured depictions. This is indicative of a tendency for AI models to rely on superficial perceptions and cultural clichés, which can result in representations that simplify or misrepresent the complexity of religious practices and symbols. Such an approach can not only misinterpret religious traditions but also reinforce prejudices and stereotypes in public perception of certain beliefs.  

\begin{figure}[htb!]
  \centering
  \includegraphics[width=0.99\textwidth]{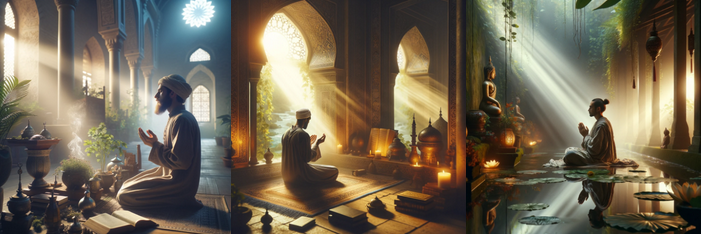}
  \caption{Religious bias. Some religious elements are more common than others, also happening with different religions, some of them misrepresented.}
  \label{fig:religious_bias}
\end{figure}

Additional research, which explores prejudices and stereotypes towards Muslims \cite{andreu2021ellos}, is a clear example of how the perpetuation of stereotypical images by the media and education can negatively influence the formation of identities and perceptions. This study brings home the importance of addressing religious bias in AI to foster more inclusive and respectful portrayals, highlighting the need to develop technologies that avoid oversimplification and promote a more qualified understanding of religious diversity. We can see an example of such a bias in Figure \ref{fig:religious_bias}.

\subsubsection{Sexual Orientation Bias}
Sexual orientation bias reveals a worrying tendency to overshadow the LGBTQ+ collective, unintentionally limiting their representation in diverse scenarios. This omission not only minimizes sexual diversity in AI-generated digital media, but also contributes to the perpetuation of heteronormative rules, excluding other forms of love expression and sexual identity. 

\begin{figure}[htb!]
  \centering
  \includegraphics[width=0.99\textwidth]{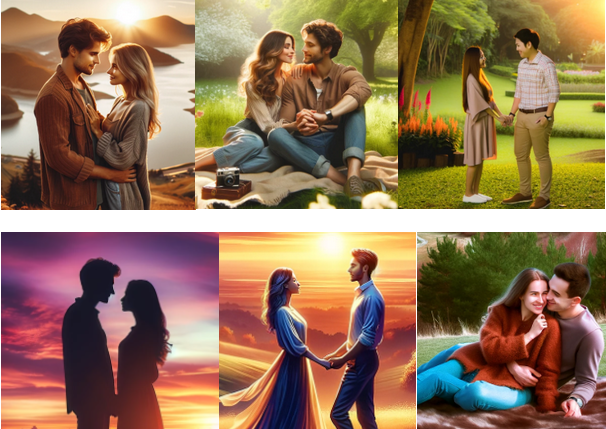}
  \caption{Sexual orientation bias. When prompted for a couple, their sexual orientation is heterosexual, apparently always painting a man and a woman, misrepresenting the LGBTQ+ collective.}
  \label{fig:sexual_bias}
\end{figure}

A relevant study in this area highlights discriminatory attitudes and the incidence of bullying based on sexual orientation, gender identity and expression (SOGIE) in educational centers \cite{zunino2020acoso}, which points to a deep-rooted social problem that goes beyond the boundaries of the classroom. The authors emphasize that bullying on the basis of SOGIE is not limited to an isolated incident but reflects broader prejudices and stereotypes in society. We can see an example of such a bias in Figure \ref{fig:sexual_bias}.

\subsection{Socioeconomic Biases}
Socioeconomic biases in artificial intelligence arise from unbalanced data sets and algorithms that omit the diversity of socioeconomic realities, unintentionally perpetuating differences in the representation of individuals. This omission leads to systems that reinforce biases and stereotypes, affecting fairness in the treatment of people from different socioeconomic backgrounds. These disparities are particularly notable in digital health technologies \cite{nazer2023bias}, highlighting the critical importance of integrating inclusive and equitable practices into all areas of AI application. 

\subsubsection{Study Bias}
The education bias emphasizes how AI models can incorrectly associate educational background with success and happiness. By this, it reflects and perpetuates social stereotypes that value college education as the main path to success, ignoring other forms of achievement and social contribution. Addressing this bias involves adjusting data sets and algorithms to promote a more equitable and realistic view of human potential, regardless of educational level, thus avoiding reinforcing prejudices and stigmatizations in society. 

\subsubsection{Income Bias}
Income bias explains how socioeconomic differences are reflected in digital representations, again suggesting a correlation between economic status and happiness. This bias can lead to an oversimplification and often erroneous simplification of reality, where people of higher economic status are portrayed in contexts of greater happiness and well-being. This bias not only reinforces harmful stereotypes about wealth and happiness but also ignores the complexity of human experiences across different socioeconomic levels. 

A relevant study in this matter investigates the impact of wealth visibility in social networks on observers' life satisfaction. The findings suggest that constant exposure to images that associate wealth with happiness may distort individuals' perception of what contributes to a satisfying life, instigating feelings of dissatisfaction and envy in those of lower economic levels. 

\subsubsection{Occupation Bias}
Occupation bias addresses how social perceptions and stereotypes influence the visual representation of various professions. This phenomenon is evidenced by the tendency of AI models to perpetuate and amplify occupational stereotypes, where certain professions are presented in ways that reinforce biases about their usefulness or social value. For example, occupations such as administrative assistance, sales and marketing, and finance are often perceived as "socially useless" \cite{walo2023bullshit}. This perception can lead to these roles being viewed in a negative or less important way, which reinforces a stereotypical and unbalanced image of these occupations in society. 

\subsection{Biological Biases}
Biological biases, in the context of artificial intelligence, refer to distortions in the representation of physical and biological characteristics, such as gender, age, or physical abilities. These biases arise when AI image generation models replicate existing biases in training datasets, affecting how people are perceived and treated on different biological characteristics. 

\subsubsection{Disability Bias}
Disability bias in AI image generation appears as a narrow and often discriminatory view towards people with disabilities. This bias is not only manifested in the lack of representation of these people in visual media, but also in the tendency to approach disability from a disadvantage perspective, ignoring the diversity and capabilities of this group. 

Nowadays, ableism is a thought that unfortunately is quite deeply rooted in society. In this research, it is highlighted how microaggressions towards people with disabilities are manifested, often minimized or ignored in the digital and social environment. Analyzing experiences shared on the social network Twitter, under the hashtag "\#MeCripple", the study indicates that many of the aggressions recorded come from contexts already analyzed in previous studies, which evidences the continuity of ableism in society. 

\begin{figure}[htb!]
  \centering
  \includegraphics[width=0.99\textwidth]{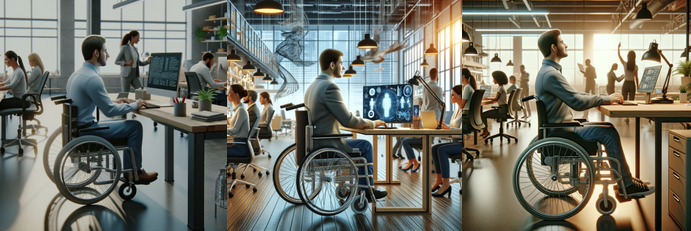}
  \caption{Disability bias. Not all are motor disabilities, which are very commonly created by the generative AI.}
  \label{fig:disability_bias}
\end{figure}

This study emphasizes the importance of creating inclusive digital spaces that allow people with disabilities to share their experiences. Extrapolating from these findings, we again note the urgency of addressing this bias in the generation of AI images to achieve a system where the diversity of all people is respected and celebrated. We can see an example of such a bias in Figure \ref{fig:disability_bias}.

\subsubsection{Gender Bias}
Gender bias covers all aspects of representation, from professions and activities to ideals of beauty and success, reflecting and perpetuating gender stereotypes. This tendency evidences an unequal assignment of roles and qualities based on gender, where, for example, beauty is predominantly associated with women and economic or professional success with men. This bias not only limits the diversity of gender representations in the digital space, but also contributes to the fulfillment of gender-based expectations, which are potentially harmful. 

\begin{figure}[htb!]
  \centering
  \includegraphics[width=0.99\textwidth]{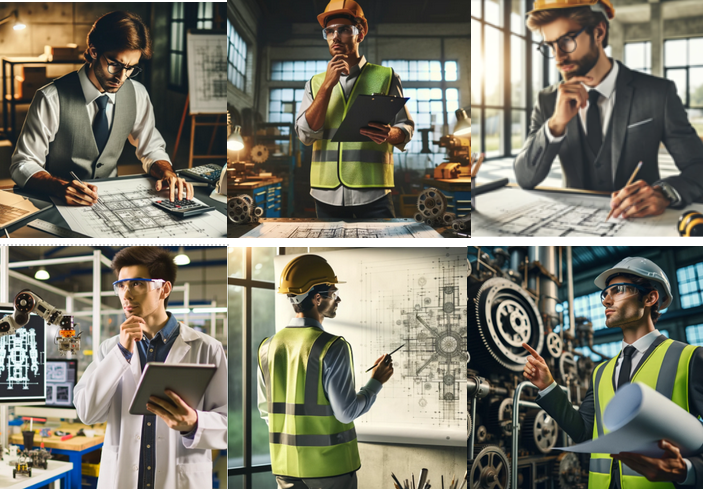}
  \caption{Gender bias on the professions of the people generated by AI, images generated asking for an engineer.}
  \label{fig:gender_bias}
\end{figure}

Research on this topic \cite{orellana2020estereotipos} has illustrated how gender stereotypes and sexist attitudes are present among university students and how these behaviors impact society. The result of this research highlights the persistence of stereotypes that assign traditional roles and specific characteristics to men and women, which can be reflected in AI models if these stereotypes are not explicitly treated in the algorithm design and training process. We can see an example of such a bias in Figure \ref{fig:gender_bias}.

\subsubsection{Age Bias}
Age bias refers to the stigmatized representation of people with respect to their age, specifically, AI image generation models tend to give visibility only to young and middle-aged adults. That is, this form of bias is evidenced in the stereotypical representation of older people, who are often made invisible or shown only in activities that reinforce negative stereotypes about old age, such as dependence or lack of activity. 

\begin{figure}[htb!]
  \centering
  \includegraphics[width=0.99\textwidth]{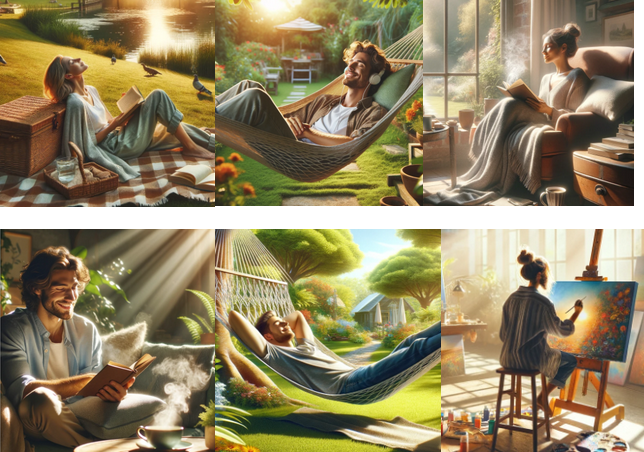}
  \caption{Age bias. When prompted for a successful person, the images represent always middle aged people.}
  \label{fig:age_bias}
\end{figure}

Ageism manifests itself through discriminations that marginalize older people \cite{cordovaedadismo}, based on the false belief that they have little contribution to offer to the community and are a burden to society. These prejudices lead to difficulties in finding employment and pension failures, making the elderly financially dependent on their families. This conceptual framework helps to understand how AI could replicate or even amplify these perceptions if the datasets used to train these systems do not account for the diversity and capabilities of people of all ages. We can see an example of such a bias in Figure \ref{fig:age_bias}.

\subsection{Demographic Biases}
Demographic biases in artificial intelligence explain how geographic origin shapes the responses of AI systems, tending to favor specific groups or stereotypical races. These disparities are due to training performed on datasets that do not capture global diversity \cite{ferrante2021inteligencia}, requiring a meticulous approach to inclusion to avoid amplifying existing biases in AI applications. 

\subsubsection{Racial Bias}
Racial bias captures the propensity to perpetuate pre-existing racial stereotypes by favoring or marginalizing certain racial groups based on data sets that are predominantly monocultural and reflect specific perspectives and biases. 

\begin{figure}[htb!]
  \centering
  \includegraphics[width=0.99\textwidth]{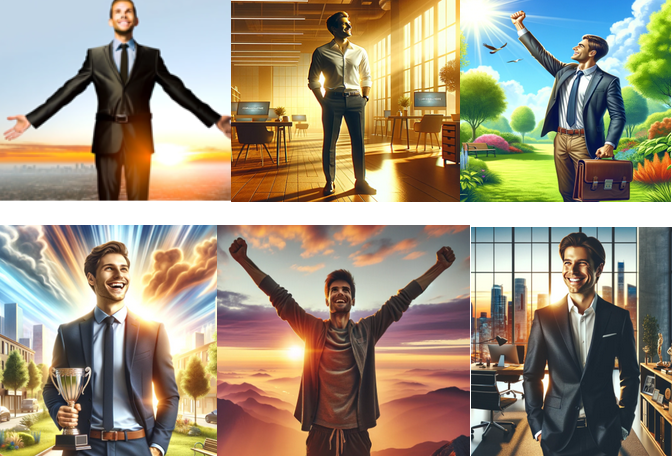}
  \caption{Racial bias that must be controlled to not misrepresent any collective.}
  \label{fig:racial_bias}
\end{figure}

As Tina Cheuk explains in her study on the use of machine learning in science assessments, algorithms can perpetuate structural inequities by favoring linguistic and conceptual norms that align with Western and White norms \cite{cheuk2021can}. The results show how students from racially minoritized groups and people with English as a second language, often face situations of disadvantage because of these biases encoded in AI. This bias not only distorts the social and cultural reality of racially diversified groups, but also reinforces racial barriers to social perception and inclusion. We can see an example of such a bias in Figure \ref{fig:racial_bias}.

\subsubsection{Locality Bias}
Locality bias reflects how geographic location affects the representation of cultures, people and traditions in the images generated by AI systems. This bias manifest itself in the tendency to stereotype and sometimes caricature the traditions and festivities of different localities, creating a sharp division between what is considered characteristic of urban and rural areas, and perpetuating social stigmas associated with these environments. The impact of this bias is considerable, as it can influence how different localities and their inhabitants are perceived and valued, leading to policies and decisions that may favor some areas over others. 

Finally, we would like to add more biases that also tend to appear but that we have not included as they are special cases of the preivously mentioned biases. Some examples are the weight of the people, being fat people misrepresented, men having beard or long hair vs bold men, among others.

\section{Why biases emerge in Generative AI content}
To understand why these models incur into bias it is first of all necessary to know that they are basically huge statistical models implemented smartly on computers. In this section, we do not care about the particular model architecture, as they in constant evolution, but in the abstract details that explain why these biases are encoded and in particular solutions that may mitigate this effect. 

Concretely, these models they can be seen  as a conditional probability distribution $p(\mathbf{Y}|\mathbf{X}, \mathbf{\theta})$ that we can sample, where $\mathbf{Y}$ is the generated image random variable, $\mathbf{X}$ is the prompt given by the user and $\mathbf{\theta}$ is a huge set of parameters that encode the patterns commonly found on the dataset of images or multimodal data $\mathcal{D}$. Depending on the prompt inserted by the user the model will condition itself to sample data from its conditional probability distribution according to its parameters $p(\mathbf{Y}|\mathbf{X}, \mathbf{\theta})$. It is in those parameters when the biases are encoded, which are transferred through the training algorithm of the model by used image dataset $\mathcal{D}$. 

The values of the parameters of these models are estimated to minimize a loss function $\mathbf{\theta}^\star = \arg\min_\Theta \mathcal{L}(\mathcal{D}, \mathbf{\theta})$ that can possibly be the error performed by the model on a benchmark or any other estimator via an optimization algorithm. Hence, if the set of images is biased according to any of the categories defined in the previous section, the algorithm does not count this fact into account in its training. For example, if regarding the religious variable, some religion is not represented into the images, then, the model parameters will not encode that particular religion and hence the generated content will not include that religion, being a bias in the generated content of the model that comes from the dataset $\mathcal{D}$. In order to detect this source of bias, we propose to design test batteries for all the biases mentioned in the test and make hypothesis statistical testing to verify whether the distributions of vulnerable variables are not biased against any particular value. If the generated content is biased, then, a potential solution is to augment the dataset $\mathcal{D}$ to include values, photos in this case, of the vulnerable variable to remove the bias present in the dataset.

Another potential solution to this bias can be implemented in the training algorithm of a model through a regularizer $r(\theta)$. In this sense, we can interpret that the training algorithm of the model can also be biased if it is not configured to minimize the potential bias of the model coming from the data. In order to solve this issue, we can use a regularizer, that basically penalizes the loss function according to some criteria to make the optimization algorithm also aware of something more than rather minimizing the prediction error or the estimator used to estimate the generalization error. If we penalize values of the parameters that incur in some biases, then, the loss function surface will change towards solutions that represent a tradeoff between performance and unbiased solutions. Through this process and via augmenting the data, we could estimate fairer models. It is obvious that, in performance terms, the model will be outperformed by a biased model, that simple encodes the biased information coming from the dataset, but, in the other hand, the obtained model is a good compromise between fairness and performance.

\section{Discussion}

The discussion on algorithmic biases in artificial intelligence has gained increasing importance in the academic and technological community, evidencing that these problems are no longer going unnoticed and are being actively addressed. Recent researchers have identified and explained how these biases manifest themselves, especially in the fields of gender and race, which are critical because of their social impact and the inequalities they can perpetuate. 

A relevant study in this area is the one conducted on Stability AI's Stable Diffusion model \cite{chauhan2024identifying}, which highlighted how gender and race biases can influence the representation of specific actions and professions. For example, this model tended to generate images of African American men playing basketball and women in nursing roles, reflecting stereotypes rooted in society. 

On the other hand, we can find a general analysis of text-to-image generation models \cite{friedrich2023fair} that outlines the importance of dealing with biases in training datasets. These models, if not properly managed, have the potential to perpetuate or even exaggerate existing biases, making it urgent to develop techniques that promote fairer and more equitable representations. 

Regarding racial biases within AI, Enzo Ferrante's study \cite{ferrante2021inteligencia} examines how facial recognition systems have been shown to perform unequally according to gender and race, being generally more effective with white male faces than with women of color or people from other racial minority groups. This phenomenon highlights the need to design AI systems that are truly inclusive and capable of treating all people equally, regardless of race or gender. 

These examples illustrate that, although considerable attention has been focused on gender and race biases, these are not the only types of biases that can affect AI systems. The taxonomy of biases explored earlier in this paper shows that there is a wide range of potential biases that also require attention and appropriate solutions to avoid perpetuating stigma and discrimination in AI image generation. 

In response to these challenges, various technical and non-technical solutions have been implemented to mitigate biases in AI models, particularly in text generation systems. According to "Improving Fairness in Deep AI" \cite{tan2020improving}, advanced techniques such as adjustments to the semantic distribution in the current space of generative networks have been adopted to improve fairness without retraining. These techniques, however, have been mainly applied in text generation, while their application in image generation has not been as widespread, leaving an important field for future research and applications.

In addition, "On Explaining Unfairness: An Overview" \cite{fragkathoulas2024explaining} highlights the importance of inclusion policies and AI ethics training. These non-technical measures complement algorithmic adjustments by promoting critical reflection on the models used and ensuring that they reflect broader diversity and representativeness. While current efforts are significant, especially in the area of gender and race bias, broader consideration of the various types of biases that can affect AI systems, as explored in this work, is required to avoid perpetuating stigma and discrimination. 

On the other hand, the legal framework in which we find ourselves must be considered. An analysis conducted by Maria José Santos González \cite{gonzalez2017regulacion} reveals that current policies to battle bias in artificial intelligence focus on transparency and accountability through bias audits and the inclusion of diversified data. These regulations, while crucial, face challenges such as defining universal standards of fairness and industry resistance to practices that increase costs or complicate development. Despite these efforts, the long-term effectiveness of these policies has yet to be fully determined, suggesting the need for continuous revision to adapt to a rapidly evolving technological field \cite{gonzalez2017regulacion}. 

Despite significant progress in the development of policies and regulations to mitigate bias in artificial intelligence, notable challenges persist that limit the effectiveness of current solutions. Existing regulations, while seeking to establish a framework of transparency and accountability, often fail to cover the complexity and rapidly changing dynamics of AI technologies. A particularly critical area is the effective enforcement of these policies, which face obstacles such as lack of resources for comprehensive audits and variability in the interpretation of what constitutes unfair outputs. In addition, current measures tend to focus predominantly on specific sectors such as facial recognition and text generation, leaving areas such as image generation with less detailed oversight and regulation.  

\section{Conclusions and Further Work}

While this study explores biases in AI image generation, it is necessary to acknowledge inherent limitations that could condition the interpretations of our findings. Specifically, the tool used may not include the full spectrum of biases present in other models. Consequently, we encourage future research to extend beyond this study, exploring a broader range of models to validate or challenge our further findings. Nonetheless, this analysis is crucial for leading the way towards a more ethical and equitable AI technologies.

Throughout this study, we have highlighted not only the gender and race biases in AI image generation, which receive the most attention, but also a broader range of socioeconomic, cultural and biological biases that influence these technologies. The taxonomy developed offers an essential tool for AI developers and users, providing a framework for identifying and mitigating potential biases in their systems. 

This investigation, significantly advances our understanding of biases in generative imaging, demonstrating the critical need for applying and adapting existing bias mitigation strategies more comprehensively within this field. While we have made advances in addressing biases traditionally observed in text generation, the unique challenge posed by image generation demand specific attention, as these biases can have a visual impact and deep cultural consequences. Therefore, this study not only highlights these issues but also sets a foundation for improving these strategies to ensure that they are effectively adapted to the complexities of image-based AI systems. 

Furthermore, it is crucial that policies and regulations evolve at the same path as AI technologies, adjusting to the challenges presented by new developments. Creating a more robust and dynamic regulatory framework that facilitates bias audits and promotes greater transparency in AI development processes will be essential to ensure that future advances in AI imaging contribute positively to society, respecting diversity and fostering inclusion. 

In light of these considerations, the AI community could work closely with marginalized groups to ensure that technological advances reflect and respect human diversity. Only through a collaborative, multidisciplinary effort we can control the power of AI to create a more inclusive and fair future. This integrated approach will fulfill AI's potential to act as a force for universal good.

AI imaging, as seen, not only reflects but potentially amplifies existing inequalities in our society. This underscores the need for interventions in both data collection and processing as well as algorithmic design to promote fairness. While this study marks a significant step towards understanding and mitigating biases in AI image generation, it also opens the door to a deeper exploration of new methodologies. As the field evolves, ongoing research must focus on refining these methods, ensuring that AI systems not only perform fairly but also sustain the principles of diversity and inclusion. This study lays the groundwork for future investigations that aim to connect general bias mitigation strategies to their application in the complex and dynamic world of AI image generation. 

\bibliography{main}
\bibliographystyle{acm}

\end{document}